\documentclass[fleqn,usenatbib]{mnras}

\usepackage{chngcntr}
\usepackage[T1]{fontenc}
\usepackage{ae,aecompl}

\usepackage{graphicx}	
\usepackage{amsmath}	
\usepackage{amssymb}	

\title[A Lensed Quasar Discovered in OGLE]{A Gravitationally Lensed Quasar Discovered in OGLE}

\author[Z. Kostrzewa-Rutkowska et al.]
{Zuzanna Kostrzewa-Rutkowska$^{1,2,3}$\thanks{E-mail:  (z.p.kostrzewa@sron.nl)}, 
Szymon Koz{\l}owski$^{3}$,
Cameron Lemon$^{4}$, \newauthor
T. Anguita$^{5,6}$,  
J. Greiner$^{7}$,
M. W. Auger$^{4}$,
{\L}. Wyrzykowski$^{3}$,
Y. Apostolovski$^{5,6}$, \newauthor
J. Bolmer$^{7,8}$, 
A. Udalski$^{3}$, M. K. Szyma{\'n}ski$^{3}$, I. Soszy{\'n}ski$^{3}$, R. Poleski$^{3,9}$, \newauthor
P. Pietrukowicz$^{3}$, J. Skowron$^{3}$, P. Mr{\'o}z$^{3}$, K. Ulaczyk$^{3,10}$, M. Pawlak$^{3}$
\\
$^{1}$SRON, Netherlands Institute for Space Research, Sorbonnelaan 2, 3584 CA Utrecht, the Netherlands\\
$^{2}$Department of Astrophysics/IMAPP, Radboud University Nijmegen, P.O. Box 9010, 6500 GL Nijmegen, the Netherlands\\
$^{3}$Warsaw University Astronomical Observatory, Al. Ujazdowskie 4, 00-478 Warszawa, Poland\\
$^{4}$Institute of Astronomy, Madingley Road, Cambridge CB3 0HA, UK\\
$^{5}$Departamento de Ciencias Fisicas, Universidad Andres Bello, Avenida Republica 220, Santiago, Chile\\
$^{6}$Millennium Institute of Astrophysics, Chile\\
$^{7}$Max-Planck Institute for Extraterrestrial Physics, D-85748 Garching, Giessenbachstr. 1, Germany\\
$^{8}$European Southern Observatory, Alonso de Cordova 3107, Vitacura, Casilla 19001, Santiago 19, Chile\\
$^{9}$Ohio State University, Department of Astronomy, 140 West 18th Avenue, Columbus, OH 43210, USA\\
$^{10}$Department of Physics, University of Warwick, Gibbet Hill Road, Coventry CV4 7AL, UK
}

\date{Accepted XXX. Received YYY; in original form ZZZ}

\pubyear{2018}

\begin{document}
\label{firstpage}
\pagerange{\pageref{firstpage}--\pageref{lastpage}}
\maketitle

\begin{abstract}
We report the discovery of a new gravitationally lensed quasar (double) from the Optical Gravitational Lensing Experiment (OGLE)
identified inside the $\sim$670 sq. deg area encompassing the Magellanic Clouds.
The source was selected as one of $\sim$60 ``red $W1-W2$'' mid-IR objects from WISE and having a significant amount of variability 
in OGLE for both two (or more) nearby sources.
This is the first detection of a gravitational lens, where the discovery is made ``the other way around'', meaning we first measured the time delay
between the two lensed quasar images of $-132<t_{\rm AB}<-76$ days (90\% CL), with the median $t_{\rm AB}\approx-102$ days (in the observer frame), 
and where the fainter image B lags image A.
The system consists of the two quasar images separated by 1.5\arcsec\ on the sky, with $I\approx20.0$~mag and $I\approx19.6$~mag, respectively,
and a lensing galaxy that becomes detectable as $I \approx 21.5$~mag source, 1.0\arcsec\ from image A, after subtracting the two lensed images. Both quasar images show clear AGN broad emission lines at $z=2.16$ in the NTT spectra.
The SED fitting with the fixed source redshift provided the estimate of the lensing galaxy redshift of $z \approx 0.9 \pm 0.2$ (90\% CL), while its type is more likely to be elliptical (the SED-inferred and lens-model stellar mass is more likely present in ellipticals) than spiral (preferred redshift by the lens model).

\end{abstract}




\section{Introduction}

Gravitational lensing is a powerful tool; acting as a natural telescope it can potentially allow astronomical observations to reach smaller and fainter objects than would otherwise be observable (\citealt{1992grle.book.....S,2006glsw.conf....1S}). In strong lensing a massive foreground galaxy or cluster deflects the light from a background object, resulting in multiple images of the distant source, that is typically magnified by a factor of 10. This provides a wide range of applications of the lensing phenomenon. Lensing analyses support fundamental open questions in astrophysics and cosmology, and have been used to investigate the dark matter density profiles of lensing galaxies (e.g., \citealt{2002ApJ...580..685S,2009ApJ...697.1892P,2014MNRAS.442.2017V,2015ApJ...800...94S}), substructures (e.g., \citealt{2002ApJ...572...25D,2004ApJ...610...69K,2010MNRAS.408.1969V}), the evolution of galaxies (e.g., \citealt{2006ApJ...649..599K,2014MNRAS.441.3238K,2017MNRAS.465.3185O}) and the properties of dark energy (e.g., \citealt{1998ApJ...502...63L,2000MPLA...15.1023Z,2010ApJ...711..201S,2012MNRAS.424.2864C}), just to mention a few.

Strong gravitationally lensed quasars with measured time delays between the multiple images provide a straightforward and competitive approach to determining the Hubble constant, entirely independent of the local distance ladder, and allow us to study the expansion history of the Universe. The time delay method was proposed by \cite{1964MNRAS.128..307R}, 15 years before the discovery of the first strong lensing system, Q$0957+561$, in 1979 (\citealt{1979Natur.279..381W}). Time delays in gravitationally lensed quasar light curves of different images depend on the gravitational potential of the lensing galaxy and the cosmological distances between the observer, the lens, and the source. In the past few years there have been many attempts to measure the Hubble constant and other cosmological parameters with increasing precision and accuracy from time delays (\citealt{2014ApJ...788L..35S,2017MNRAS.468.2590S}).
The time delay studies require not only detailed models for the mass profile of the main lens but also taking into account any matter structures along the line of sight (\citealt{2017MNRAS.467.4220R}).
Also recently, long time monitored systems were used to determine the Hubble constant with a precision below 5 per cent (\citealt{2017MNRAS.465.4914B}). 

Furthermore, in multiple quasar images, we may observe the microlensing effect caused by the stars of the lens (e.g., \citealt{1990ApJ...358L..33W,1991AJ....102..864W,2002ApJ...580..685S}). This provides stringent constraints on the morphologies of the neighbourhood of black holes in lensed quasars, e.g. kinematics and geometric structure in quasar broad emission-line regions (\citealt{2014A&A...565L..11B,2016A&A...592A..23B,2013ApJ...764..160G}). Detailed studies of the accretion disks showed that their sizes are robustly larger than expectations from the thin disk theory and the temperature profile seems to be steeper than expected (\citealt{2007MNRAS.381.1591B,2008MNRAS.391.1955B,2015ApJ...806..258M,2014ApJ...783...47J,2015ApJ...806..251J}), however, opposite results have also been reported (\citealt{2008ApJ...673...34P}). Because the microlensing effect is also sensitive to the fraction of mass enclosed in stars to dark matter halo mass close to lensed images, at the same time we obtain the stellar composition of the lens galaxies (\citealt{2011ApJ...731...71B,2014ApJ...793...96S,2015ApJ...799..149J}). While microlensing of lensed quasar images has numerous applications to study the AGN physics, on the other hand, it affects the light curves of quasar images, producing problems in the accurate time delay estimation, and in turn introducing biases in the characterisation of lens systems (\citealt{2017arXiv170701908T}). However, recently several algorithms were applied to take into account microlensing in the time delay measurements (e.g. \citealt{2017MNRAS.465.4914B}).

There are less than two hundred strongly lensed quasars known to date, because (1) they are intrinsically very rare objects 
and (2) they are discovered by various techniques, each having a number of biases.
Considering that we search for lensed sources blended with the light of the lens galaxy, the problem of the deflector's flux contribution impedes the lensed quasar hunt.
Systematic searches of gravitationally lensed quasars have been conducted within several surveys, for example, by the Sloan Digital Sky Survey (SDSS) and the Dark Energy Survey (DES). Recently, several techniques were explored to search for lensed quasar systems, e.g. identification of spectroscopically confirmed quasars and searching for nearby objects of similar colours (large separation systems) or spectroscopically confirmed quasars with extended morphology (small separation systems) (\citealt{2006AJ....132..999O,2008AJ....135..496I,2016MNRAS.456.1595M}), selection of lensed quasars using an image subtraction method as they are time variable sources \citep{2006ApJ...637L..73K,2009ApJ...698..428L}, data mining and mashine learning techniques with a combination of colour cuts and model-based selection \citep{2015MNRAS.454.1260A,2017AJ....153..219S}, Gaussian mixture models of colours \citep{2017MNRAS.465.4325O}, identifying candidate systems of red galaxies with multiple neighbouring blue objects \citep{2017ApJ...838L..15L}, and variability studies of quasars \citep{2017ApJ...844...90B}.

One of the primary goals of the Optical Gravitational Lensing Experiment (OGLE) -- a long-term sky-variability survey (\citealt{2015AcA....65....1U}) -- has been monitoring of selected gravitational lenses. 
Since 1997, OGLE has performed a continuous, long-term monitoring of the gravitational lens QSO 2237+0305 \citep{2006AcA....56..293U}, where the resulting homogeneous OGLE dataset is the most extensive photometric coverage of this source, spanning now two decades and showing numerous unique microlensing effects in this extraordinary object. For the second monitored system, HE 1104--1805, the time delay of $157 \pm 21$ days between the light curves of the two images was determined using the photometric data collected between years 1997 and 2002 \citep{2003AcA....53..229W}.

In this paper, we present the discovery of the first lensed quasar found in the OGLE dataset covering the Magellanic System, the area that has been regularly monitored since 2010. Inside the observing area of about 670 square degrees around the Magellanic Clouds to the photometric depth of $I < 21$~mag, statistically there should exist about 10 lensed QSO systems, both doubles and quads (\citealt{2010MNRAS.405.2579O}). Finding all of them in a search is difficult as the detection efficiency is rarely high, mainly due to the fact that  quasar variability amplitude becomes buried in the photometric noise for objects fainter than $I\approx20$~mag.
This is true in our search, where we combined the OGLE optical data with the mid-IR Wide-field Infrared Survey Explorer (WISE) data, and searched for the
intrinsic optical quasar variability (in particular time lags) in OGLE counterparts to ``red'' mid-IR sources, without prior spectroscopic analysis. The details of our search and results will be presented in
subsequent sections.

This paper is organized as follows. In Section 2 we introduce our data sample and present selection method, then describe the lensed quasar candidate (that we subsequently confirm) in Section 3. In Section 4, we discuss results of modelling and lens and source properties. The paper is concluded in Section 5.
Throughout this paper we assume a flat $\Lambda$-Cold Dark-Matter ($\Lambda$CDM) concordance cosmological model of the Universe with parameters $\Omega_\Lambda = 0.7$, $\Omega_\mathrm{M} = 0.3$ and $H_0 = 70\mathrm{~km~s^{-1}~Mpc^{-1}}$, $h = 0.70$.

\section{The OGLE survey and lensed quasars detection strategy}
\label{sec:search}

\subsection{The OGLE Variability Survey}

The key data used in this work were collected during the fourth phase of the OGLE project (hereafter OGLE-IV), between the years 2010 and 2017, with the 1.3-m Warsaw telescope located at Las Campanas Observatory, Chile, operated by the
Carnegie Institution for Science. The telescope is equipped with a 32 chip mosaic CCD camera, with a total field of view of 1.4 square degrees (\citealt{2015AcA....65....1U}). The entire area of the Magellanic Clouds region is covered by 475 OGLE-IV fields that amounts to about 670 square degrees. Only two filters $I$- and $V$-band have been used by OGLE-IV. About 90 per cent observations were carried out in the $I$-band with magnitude range from 13 to 21.5 mag. The number of collected epochs varies between 100 and 700, and observations of a given field are repeated every 2--4 days. The photometry was obtained using the difference image analysis (DIA) method, as implemented by \citet{2000AcA....50..421W}.

\subsection{Lensed quasar candidates}

Since 2010, The Magellanic Quasars Survey (MQS) has increased the number of quasars known behind the Magellanic Clouds by almost an order of magnitude using the OGLE-III data (\citealt{2013ApJ...775...92K}). It was then a natural consequence to perform a similar extensive search of quasar candidates in the increased OGLE-IV area of the Magellanic System (Koz{\l}owski et al. in prep.). 

The primary sample of quasar candidates was obtained by locating all objects in the WISE survey (\citealt{2010AJ....140.1868W,2014yCat.2328....0C}) fulfilling any of the mid-IR colours criteria for quasars proposed by \cite{2012ApJ...753...30S} and \cite{2013ApJ...772...26A}, then cross-matching them with objects in the OGLE database, and finally performing a variability analysis of the OGLE objects, as a result isolating a sample of quasar candidates ($\sim$15000 objects).

We used this sample of quasar candidates in further identification of lensed quasar candidates. First, we investigated their environments within a radius of $r<6$ arcsec (approximately W1 WISE FWHM) to identify neighbouring objects in the OGLE database. Of the 15000 sources, 63 objects turned out to be ``our early'' lens candidates based on the similar rms. in the light curves. Next, for each such a lens candidate (now two or more neighbouring objects), we visually compared both their colour and variability properties. The highest probability candidates were required to have similar variability patterns in both images (and so plausibly the time delay) as reported by the {\sc Javelin}\footnote{\tt https://bitbucket.org/nye17/javelin} software (\citealt{2011ApJ...735...80Z,2013ApJ...765..106Z,2016ApJ...819..122Z}). We also used a dedicated difference image analysis (to prevent ``ghost'' variability in the neighbour caused by the true variable object -- we created two images, where we stacked a number of images when the quasar is bright and when it is faint, and then subtracted them to check if the residua due to flux change appear on both locations).
Three most promising candidates were observed spectroscopically with the New Technology Telescope (NTT), of which only one turned out to be a genuine lensing system (the other two are unlensed quasar+star pairs, presented in Figure~\ref{fig:contam}).
The false positive rate might be lower by, first, a visual inspection of both the light curves and difference images (ghosts), and then by the spectral energy distribution (SED) fitting. 
This method allowed for a straightforward identification of the lensed quasar candidate system without extensive spectroscopic observations. 

\section{The Lens System}

\begin{figure}
    \includegraphics[width=\columnwidth]{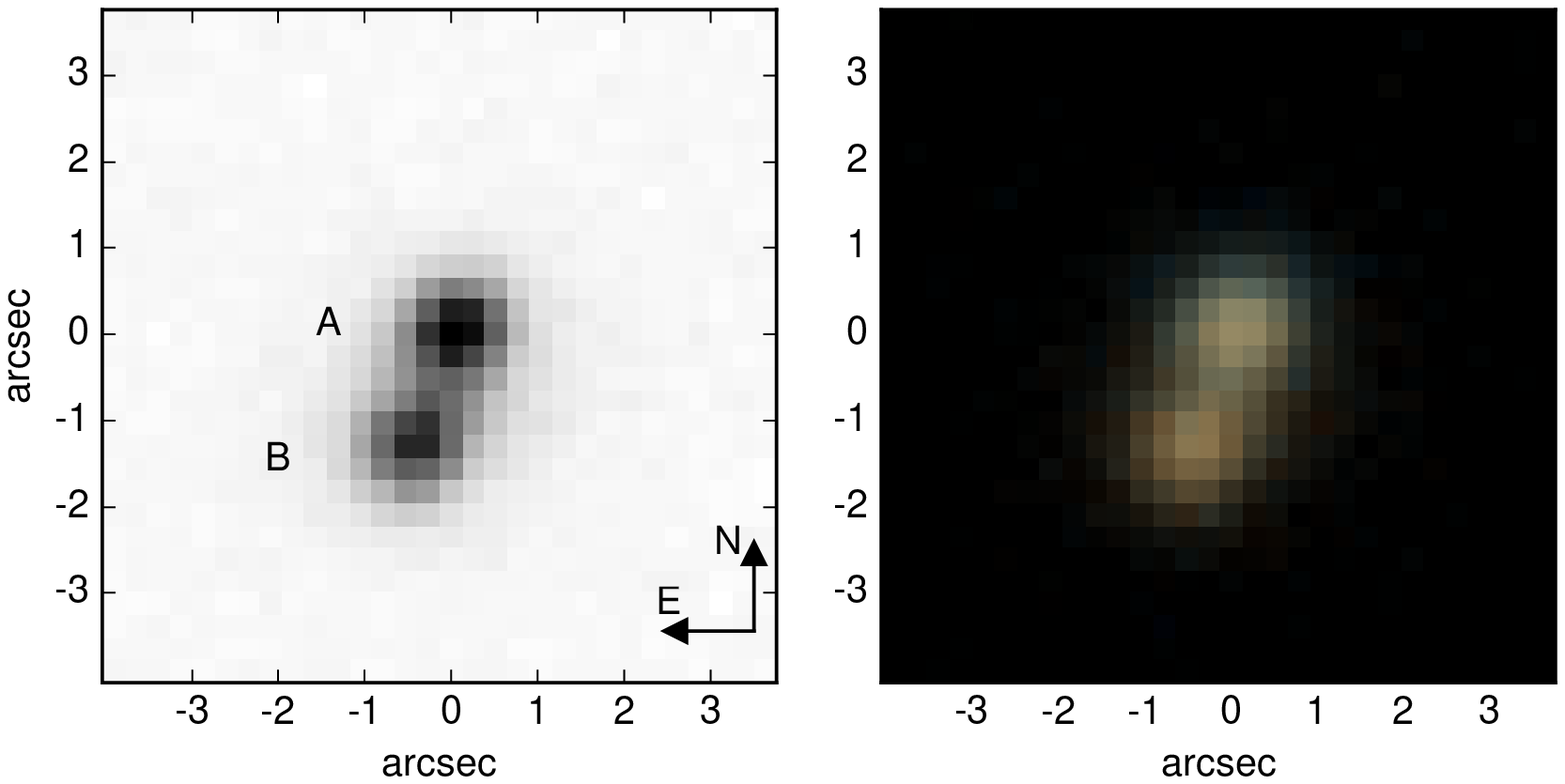}
    \includegraphics[width=\columnwidth]{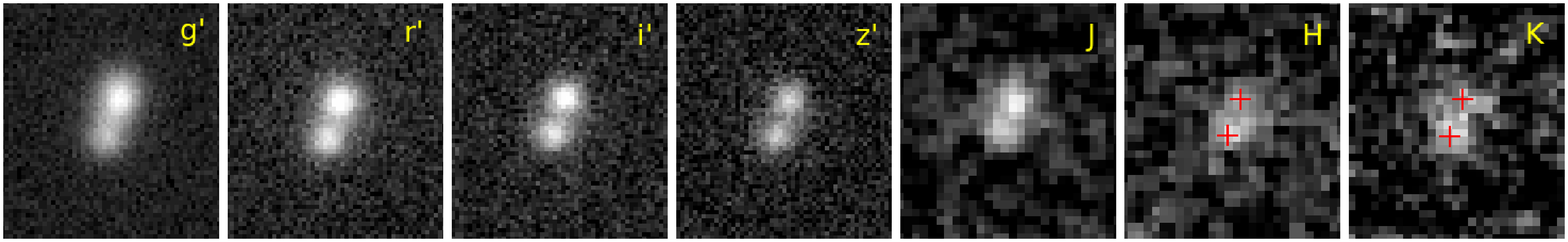}
    \caption{The OGLE lensed quasar as seen in OGLE (top) and GROND (bottom) is shown. The two lensed quasar images are marked in the top left panel, where A (B) is the brighter (fainter) image. They are separated by 1.5\arcsec\ and are readily resolvable in the best-seeing OGLE frames of 0.7\arcsec. The top right panel shows the colour composition of $I$- and $V$-band frames. The bottom row presents $g'r'i'z'JHK$ GROND observations (from left to right). The lens galaxy becomes visible as a blend to image B in the infrared data. All panels cover approximately $8\arcsec\times8\arcsec$. North is up and east is to the left. }
    \label{fig:findch}
\end{figure}

Fulfilling the requirements from Section~\ref{sec:search}, the candidate considered in this paper had to have ``red'' WISE colours ($W1-W2>0.8$~mag from \citealt{2012ApJ...753...30S} or a combination of $W1-W2$ and $W2$ data from \citealt{2013ApJ...772...26A}) but also two optical counterparts well separated on the sky, both showing similar photometric variability, albeit allowed to be shifted in time. In Figure \ref{fig:findch}, we present the postage-stamp cutouts for the candidate images A and B, at (RA, Decl.) $=(02{:}18{:}12.34$, ${-}73{:}35{:}35.8$) and ($02{:}18{:}12.47$, ${-}73{:}35{:}37.2$), respectively.

\subsection{Spectroscopic confirmation}
\label{sec:spectroscopy}

\begin{figure}
	\includegraphics[width=\columnwidth]{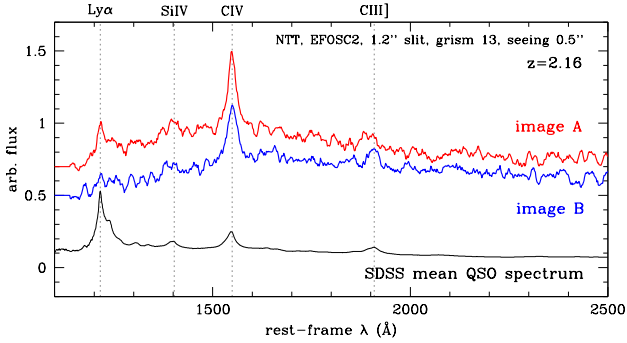}
    \caption{NTT spectra of the two images (A in red, B in blue) of the lensed quasar. 
    For comparison, in black, we also show a composite AGN spectrum from \protect\cite{2001AJ....122..549V}. 
    The source redshift is $z_S=2.16$. Four common quasar emission lines are marked with the 
    vertical dotted lines and labelled at the top.}
    \label{fig:spec}
\end{figure}

We obtained a low resolution spectrum of our candidate with the ESO Faint Object Spectrograph and Camera (EFOSC2) on the 3.58m Ritchey-Chr{\'e}tien New Technology Telescope (NTT) on 2016 September 27. The spectrum of 300s was taken with a 1.2\arcsec slit, aligned to contain both objects, and using the 13 grism, which gives a wavelength range of 3685 to 9315\AA. The data were binned by factors of 2 along the slit and in the spectral direction. This setup yields a dispersion of 5.54\AA\ per pixel and a FWHM of 21.24\AA. The seeing ranged between 0.5--0.6\arcsec. The pair of images in the candidate system was completely resolved. The 2D spectrum was reduced by fitting two Gaussians and extracting 1D spectrum. The confirmation spectrum is shown in Figure~\ref{fig:spec}. Several common AGN broad emission lines (Ly$\alpha$, SiIV, CIII], and CIV) at the redshift of $z_S=2.16$ are present. We were unable to recognise any features from a potential lensing galaxy, simply because it was expected to be too faint in this wavelength range (see SED in Figure~\ref{fig:sed}). 

\subsection{Estimation of the lens redshift}

The estimate of the lens redshift comes from the SED fitting.
In order to build an SED, we combined a number of datasets from the literature and observations, totalling 20 bands from UV to mid-IR.
The near-UV (NUV) data came from Galex (\citealt{2011Ap&SS.335..161B}), while both $V$- and $I$-band magnitudes were obtained by OGLE-IV (\citealt{2015AcA....65....1U}). 
We performed dedicated observations of the candidate with GROND (PI: Jochen Greiner) in $g'$, $r'$, $i'$, $z'$, $J$, $H$, $K$ filters (\citealt{2008PASP..120..405G}),
operated as an imaging instrument at the MPI/ESO 2.2m telescope in La Silla, Chile. 
The mid-IR data were obtained both from the Spitzer IRAC and MIPS detectors (C1--C4 and C5 channels; the SAGE project; \citealt{2006AJ....132.2268M,2011AJ....142..102G}) and WISE ($W1$--$W4$ bands; \citealt{2010AJ....140.1868W,2014yCat.2328....0C}) space-based telescopes, both covering the 3--24 micron range (Table~\ref{tab:SED}).
These broad-band magnitudes were converted to luminosities, as prescribed by \cite{2015AcA....65..251K}, using the standard $\Lambda$CDM cosmological model. 

\begin{figure}
    \centering
	\includegraphics[width=0.8\columnwidth]{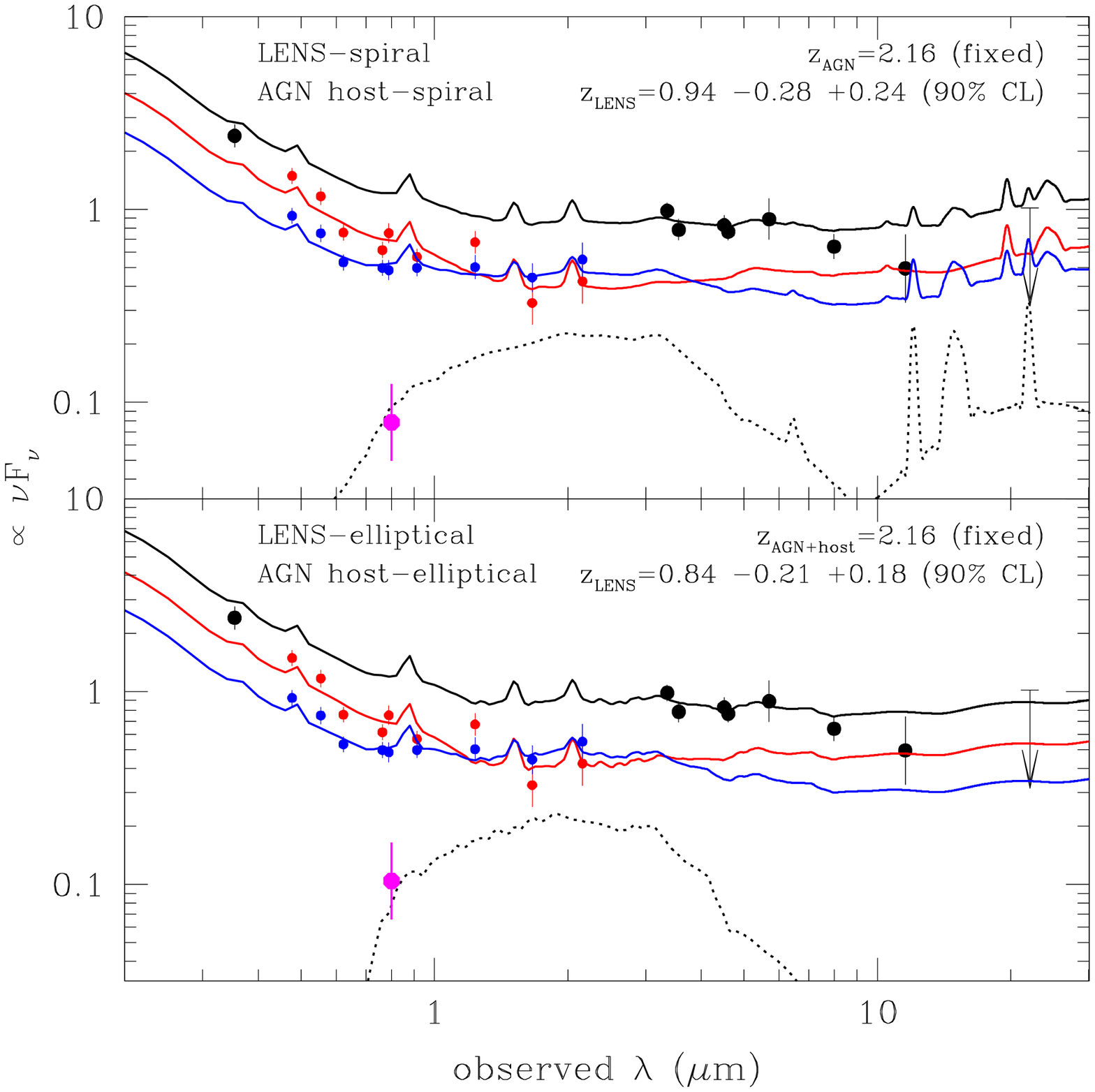}\\
    \vspace{0.3cm}
    \includegraphics[width=0.82\columnwidth]{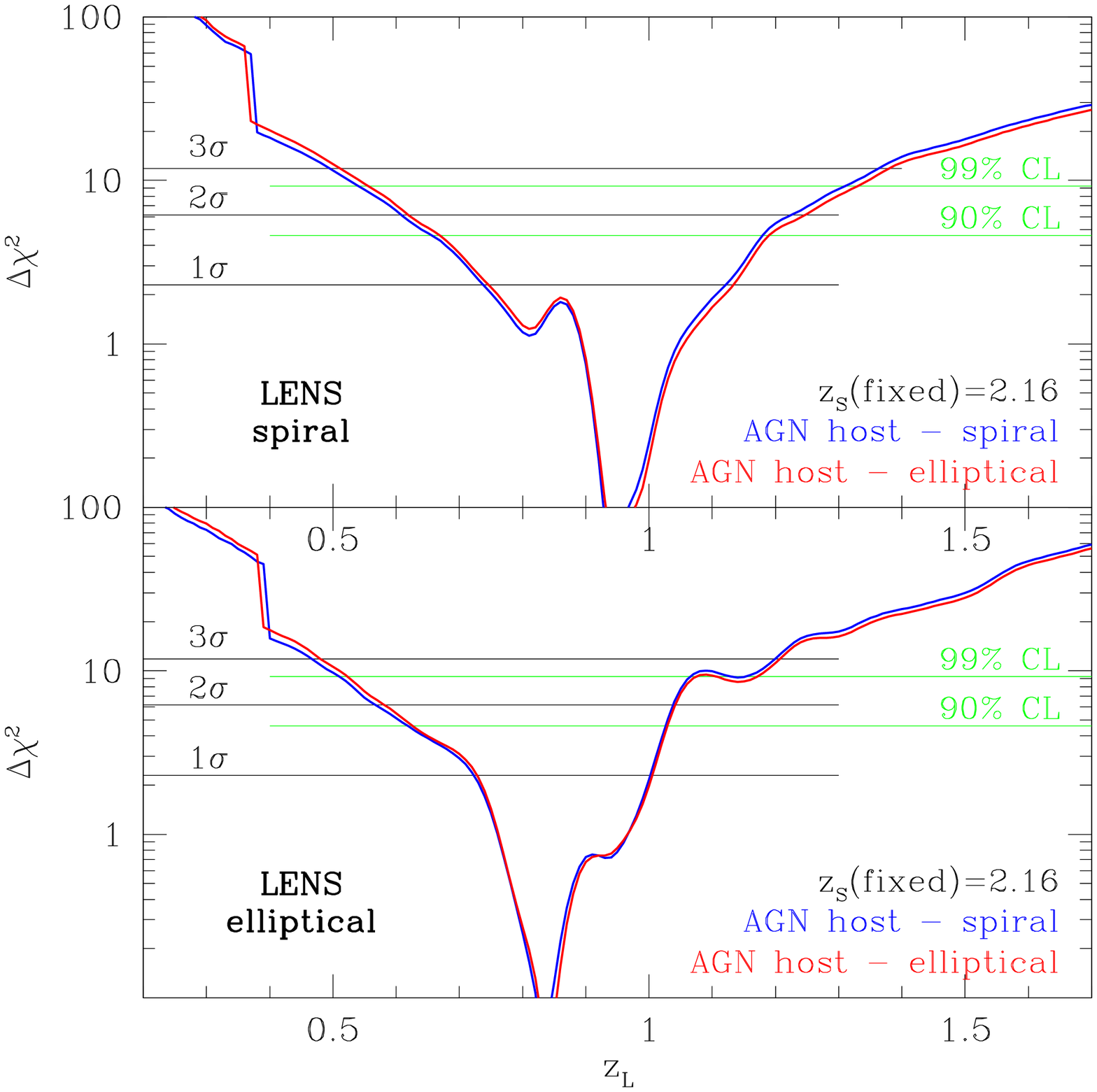}
    \caption{SED of the OGLE lens system (top) and the $\chi^2$ distribution for the lens redshift (bottom). {\it Top:} The black, red, and blue points come from the OGLE, GROND, Spitzer, and WISE surveys (Table~\ref{tab:SED}), where black is for the two images of the quasar being unresolved, while the red  and blue points are for the resolved images A and B, respectively. Image A (red) is assumed to be a sum of the AGN and its host (at $z_S=2.16$), while image B (blue) due to proximity of the lens galaxy is a sum of the AGN, its host, and the lensing galaxy. In both sub-panels, the solid black line is the best-fitting model, being the sum of the AGN and the host, and a lens galaxy (dashed line): spiral or elliptical. 
    The magenta point is the constraint on the lens magnitude in $I$-band from the PSF fitting. 
    {\it Bottom:} $\Delta\chi^2$ as a function of the lens redshift for two types of lensing galaxies and two types of host galaxies.}
    \label{fig:sed}
\end{figure}

To fit the SED (Figure~\ref{fig:sed} and Table~\ref{tab:SED}), we used the low-resolution templates of AGN and galaxies from \cite{2010ApJ...713..970A}. 
We excluded the Galex NUV point from the fitting, because of its unreasonable low value as compared to models.
The origin of this discrepancy lies in the fact that the UV flux in distant AGNs is absorbed by neutral hydrogen in the intergalactic medium, a feature that is not taken into account in both the templates used and our simplistic SED modelling. 
Both the OGLE and GROND data
have sufficiently high resolution to measure fluxes for both lensed images, while the Spitzer and WISE data show only a single object due to blending or insufficient resolution.
Whenever possible, we fitted the image A (as a sum of the AGN and the host galaxy: either spiral or elliptical) separately from the image B (a sum of the AGN, the host and the lensing galaxy that blends with the image B).
We only roughly know the $I$-band magnitude of the lens galaxy ($I\approx21.5$~mag), that we used as an additional constraint in the SED fitting. We also know the source redshift, so we fixed it in the SED fitting to $z_S=2.16$. We obtained the lens redshift of $z_L=0.94_{-0.28}^{+0.24}$ (90\% CL) for the spiral (insensitive to the AGN host type) and $z_L=0.84_{-0.21}^{+0.18}$ (90\% CL) for the elliptical galaxy (also insensitive to the host type).
Best-fitting galaxy and AGN models as well as the $\chi^2$ distributions are presented in Figure~\ref{fig:sed}. We have included the Galactic extinction model as proposed by \cite{1989ApJ...345..245C} with $R_V=3.1$, however, the results turned out to be non-physical and pointing to negative extinctions. Fixing the extinction to positive values provides a slightly lower redshift for the lens galaxy at a price of higher $\chi^2$.
From the SED fit, we can also estimate the lens galaxy stellar mass: $(2.5\pm 0.5)\times10^{11} \mathrm{M_\odot}$ for the elliptical and $(3.3\pm0.7) \times10^{11} \mathrm{M_\odot}$ for the spiral, however the latter is more than an order of magnitude too high for this type of galaxies (\citealt{2014MNRAS.440..889S}).

\begin{table}
	\centering
	\caption{Combined observational properties of the system across the electromagnetic spectrum (SED).\label{tab:SED}}
	\begin{tabular}{lcccr} 
		\hline
	filter$^{\star\dagger}$ & $\lambda$ & image & magnitude$^{\ddagger}$ &  $\propto \nu F_\nu$ \\
		\hline
 NUV (AB)    &  0.23  & AB & $21.66 \pm 0.30$ & $ 0.54 \pm  0.13$ \\ 
   u (AB)    &  0.35  & AB & $19.57 \pm 0.11$ & $ 2.41 \pm  0.23$ \\ 
   \hline
   g' (AB)$^{\star}$   &  0.48  & A  & $19.77 \pm 0.02$ & $ 1.49 \pm  0.03$ \\ 
   V (Vega)  &  0.55  & A  & $19.94 \pm 0.02$ & $ 1.17 \pm  0.01$ \\ 
   r' (AB)   &  0.62  & A  & $20.22 \pm 0.02$ & $ 0.76 \pm  0.01$ \\ 
   i' (AB)   &  0.76  & A  & $20.24 \pm 0.02$ & $ 0.62 \pm  0.01$ \\ 
   I (Vega)  &  0.79  & A  & $19.52 \pm 0.02$ & $ 0.75 \pm  0.01$ \\ 
   z' (AB)   &  0.91  & A  & $20.11 \pm 0.04$ & $ 0.57 \pm  0.01$ \\ 
   J (Vega)  &  1.24  & A  & $18.70 \pm 0.11$ & $ 0.68 \pm  0.04$ \\ 
   H (Vega)  &  1.66  & A  & $18.68 \pm 0.26$ & $ 0.33 \pm  0.10$ \\ 
   K (Vega)  &  2.16  & A  & $17.65 \pm 0.27$ & $ 0.42 \pm  0.11$ \\ 
   \hline
   g' (AB)   &  0.48  & B  & $20.28 \pm 0.02$ & $ 0.93 \pm  0.02$ \\ 
   V (Vega)  &  0.55  & B  & $20.35 \pm 0.02$ & $ 0.75 \pm  0.04$ \\ 
   r' (AB)   &  0.62  & B  & $20.58 \pm 0.02$ & $ 0.53 \pm  0.01$ \\ 
   i' (AB)   &  0.76  & B  & $20.45 \pm 0.03$ & $ 0.50 \pm  0.01$ \\ 
   I (Vega)  &  0.79  & B  & $19.93 \pm 0.02$ & $ 0.48 \pm  0.01$ \\ 
   z' (AB)   &  0.91  & B  & $20.25 \pm 0.03$ & $ 0.50 \pm  0.02$ \\ 
   J (Vega)  &  1.24  & B  & $19.00 \pm 0.12$ & $ 0.50 \pm  0.05$ \\ 
   H (Vega)  &  1.66  & B  & $18.35 \pm 0.15$ & $ 0.44 \pm  0.06$ \\ 
   K (Vega)  &  2.16  & B  & $17.37 \pm 0.20$ & $ 0.55 \pm  0.08$ \\ 
   \hline
  W1 (Vega) &  3.35  & AB & $15.42 \pm 0.04$ & $ 0.98 \pm  0.03$ \\ 
  C1 (Vega) &  3.56  & AB & $15.50 \pm 0.09$ & $ 0.79 \pm  0.06$ \\ 
  C2 (Vega) &  4.51  & AB & $14.70 \pm 0.08$ & $ 0.83 \pm  0.06$ \\ 
  W2 (Vega) &  4.60  & AB & $14.69 \pm 0.04$ & $ 0.77 \pm  0.03$ \\ 
  C3 (Vega) &  5.69  & AB & $13.88 \pm 0.12$ & $ 0.89 \pm  0.61$ \\ 
  C4 (Vega) &  7.98  & AB & $13.24 \pm 0.12$ & $ 0.64 \pm  0.07$ \\ 
  W3 (Vega) & 11.56  & AB & $12.39 \pm 0.18$ & $ 0.49 \pm  0.08$ \\ 
  W4 (Vega) & 22.09  & AB & $ < 9.41$ & $ < 1.02$ \\ 
  C5 (Vega) & 23.68  & AB & $ 9.72 \pm 0.08$ &  $ 0.62 \pm  0.16$ \\ 
	\hline
    \multicolumn{5}{l}{$^\star$C/W stands for Spitzer-IRAC(MIPS)/WISE bands.}\\
    \multicolumn{5}{l}{$^\dagger$ The photometry is provided in either AB or Vega systems.}\\
    \multicolumn{5}{l}{$^\ddagger$ The magnitudes are extinction corrected.}\\
    \multicolumn{5}{l}{$^\star$ SDSS magnitude system as desribed in \cite{1996AJ....111.1748F}}
	\end{tabular}
\end{table}

\subsection{Astrometry from OGLE}
\label{sec:astro}

\begin{figure}
    \includegraphics[width=\columnwidth]{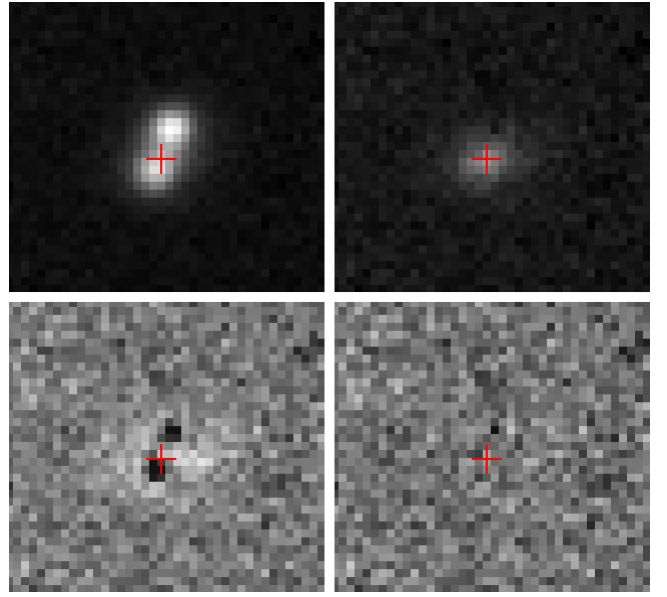}
    \caption{The results of image modelling. The OGLE $I$-band image is shown (top-left panel). It was modelled as a sum of two PSFs and a galaxy. The two lensed quasar
images obtained from this model were removed in the top-right panel, leaving only the galaxy (marked with the red crosshair in all panels).
Residuals after subtracting the best-fitting model including two PSFs only is shown in the bottom-left panel. 
It is clear such a model was insufficient to correctly describe the data. The best model included both the two PSFs and the galaxy, then the residuals
became smooth and showed no obvious additional light (bottom-right panel). Astrometry and basic parameters of this system are provided in Table~\ref{tab:system}. All panels cover $8\arcsec\times8\arcsec$, north is up, east is to the left.}
    \label{fig:galres}
\end{figure}

Basic properties of the lensing system are given in Table~\ref{tab:system} and the finding chart is shown in Figures~\ref{fig:findch} and \ref{fig:galres}.
The lensed system is clearly resolved as two point sources separated by 1.5\arcsec\ in the OGLE $I$-band imaging data. 
The image pixels were modelled to obtain the astrometry of the system. 
First, we determined the point spread function (PSF) by fitting a nearby star with a parametric Moffat profile. 
We used four different nearby stars (within $\sim$ 1 arcmin) and compared the residuals and the Moffat shape parameters. All the stars are brighter than the quasar images. The brightest star did not give a reasonably good fit  consistent with shape parameters for the other stars and it was excluded from the final fit. The Moffat $\beta$ parameter changes between fits but is $\sim$2.5. For the final lensed quasar fit we accounted for the uncertainty in the PSF, hence we allowed the PSF parameters to vary within the measured values (of the other 3 stars) and then marginalise over them.
The system was then modelled as a combination of two quasar images (two PSFs) and a galaxy profile, that was set to be a de Vaucouleurs profile convolved with the PSF profile. 
The free parameters of our model were the 2D positions of the quasar images and lensing galaxy, and the galaxy shape parameters (flattening and effective radius). 
The model parameters were determined, along with their uncertainties, via sampling with the 
Markov Chain Monte Carlo (MCMC) ensemble sampler -- {\sc emcee }  (\citealt{2013PASP..125..306F}). 

In Figure \ref{fig:galres}, we show the original data (top-left panel).
They were modelled as a sum of two PSFs and a galaxy. The top-right panel presents the lensing galaxy with $I\approx 21.5$ mag, 
that becomes visible after subtracting the two model PSFs (and not subtracting the galaxy) from the image.
Modelling of this image with two PSFs only results in bad residuals after subtracting the model from the data (bottom-left panel). 
The data are sufficiently modelled only if both the images and the galaxy are included. The residuals after subtracting such a model from the data are clean
 (bottom-right panel).
The lens light profile was modelled with de Vaucouleurs profile and we obtained the shape parameters: the effective radius of $0.55\pm0.08$ arcsec, axis ratio $0.51\pm0.07$ and position angle $101\pm4$ (in degrees measured east of north).

\begin{table*}
	\centering
	\caption{Basic properties of the OGLE lens system.\label{tab:system}}
	\begin{tabular}{lccccccr} 
		\hline
		object & RA    & Dec  & $\Delta$RA & $\Delta$ Dec. & V  & I & redshift\\
		       &       &      & (arcsec)   & (arcsec)      & (mag) & (mag) & \\ 
		\hline
		image A & $02{:}18{:}12.34$ & ${-}73{:}35{:}35.8$ &    0   & 0 & 20.09 & 19.61 & 2.16\\
		image B & $02{:}18{:}12.47$ & ${-}73{:}35{:}37.2$ & $0.54\pm0.01$   & $-1.44\pm0.01$ & 20.50 &20.02 & 2.16\\
		galaxy  & $02{:}18{:}12.42$ & ${-}73{:}35{:}36.8$ & $0.32\pm0.03$   & $-1.01\pm0.03$ & $\cdots$  & $\approx 21.5$ & $\approx 0.9$ \\
		\hline
	\end{tabular}
\end{table*}

\subsection{The OGLE photometry}
In Table~\ref{tab:LCs}, we present the $I$-band light curves (419 epochs) for the two lensed images of the quasar, along with
a basic observational log: airmass, seeing and background, enabling trimming the data from bad weather points.
Note, that the light curves are only best-calibrated to the Vega system, 
but not standard, as they lack the $(V-I)$ colour term, because there is much less data points in $V$-band than in $I$-band.
The image subtraction method typically reports underestimated error-bars that we corrected
using the following equation $\Delta I_{\rm new} = \sqrt{\left(\eta\Delta I_{\rm DIA}\right)^2+\epsilon^2}$,
where $\eta=1.558$ and $\epsilon=0.004$ (\citealt{2016AcA....66....1S}).

From the OGLE photometry of template images (now in the standard Johnson-Cousins Vega system), we find that the colours of the lensed images are $(V-I)_\mathrm{A}=0.48$~mag and $(V-I)_\mathrm{B}=0.78$~mag, so the image B is redder than A, what is also readily visible in the top-right panel of Figure~\ref{fig:findch}. This may point to two scenarios: 
(1) image B is a composite of the lensing galaxy and the lensed quasar, or
(2) image B is redder due to extinction in the lensed galaxy. If scenario (1) is the case, then we can easily estimate the $I$-band magnitudes of the two sources (assuming the $V$-band lens galaxy flux is negligible). The lensing galaxy brightness is $I=21.23$~mag and the lensed image has the brightness of $I_\mathrm{B}=20.02$ mag (Table~\ref{tab:system}). The lensing galaxy $I$-band brightness is confirmed by the measurement obtained from the PSF fitting. On the other hand, from the SED modelling, we find the ``pure'' flux ratio between images A (61\% of light of the total AB light) and B (39\%) is $\sim 1.6 $ with the formal uncertainty of $0.1$, therefore the magnitude difference between images is $\sim$0.5 mag, however, the lensing galaxy contributes significantly to the flux of image B due to its proximity. In the near-IR, the flux ratio decreases to below 1.0, where the image B appears to be brighter than A, but in fact the best SED model of the lensing galaxy peaks around 2 $\mu$m and dominates the light of image B (bottom panel of Figure~\ref{fig:findch} and top panel of Figure~\ref{fig:sed}).

\begin{table*}
	\centering
	\caption{The OGLE light curves for the two images of the lensed quasar.\label{tab:LCs}
(For the guidance purposes, we provided below the first five rows of the full light curves, that are available in its entirety from the electronic version of the journal.)}
	\begin{tabular}{lccccccr} 
		\hline
		HJD-2450000 &  image A & image A & image B & image B  & airmass & seeing & background \\
		(days) & I (mag) & $\sigma_{\rm I}$ (mag) & I (mag) & $\sigma_{\rm I}$ (mag) & & (\arcsec) & (counts) \\
		\hline
                5377.90992 & 19.801 & 0.156 & 19.708 & 0.143 & 1.528 & 1.55 &  696 \\
                5386.92763 & 19.898 & 0.080 & 19.870 & 0.080 & 1.450 & 1.12 &  253 \\
                5388.92893 & 19.846 & 0.087 & 19.942 & 0.097 & 1.441 & 1.24 &  245 \\
                5389.93796 & 20.010 & 0.125 & 19.974 & 0.122 & 1.428 & 1.61 &  336 \\
                5392.92310 & 19.975 & 0.154 & 19.875 & 0.140 & 1.435 & 2.01 &  320 \\
		\hline
	\end{tabular}
\end{table*}


\subsection{The time delay}
\label{sec:timedelay}

Time delays in gravitational lenses may be the key to understanding the lens system, including
its mass distribution, but also are probes of cosmological parameters such as the Hubble constant (e.g.,  \citealt{1986ApJ...310..568B,1988ApJ...327..693G,2002ApJ...578...25K}).

\begin{figure}
	\includegraphics[width=\columnwidth]{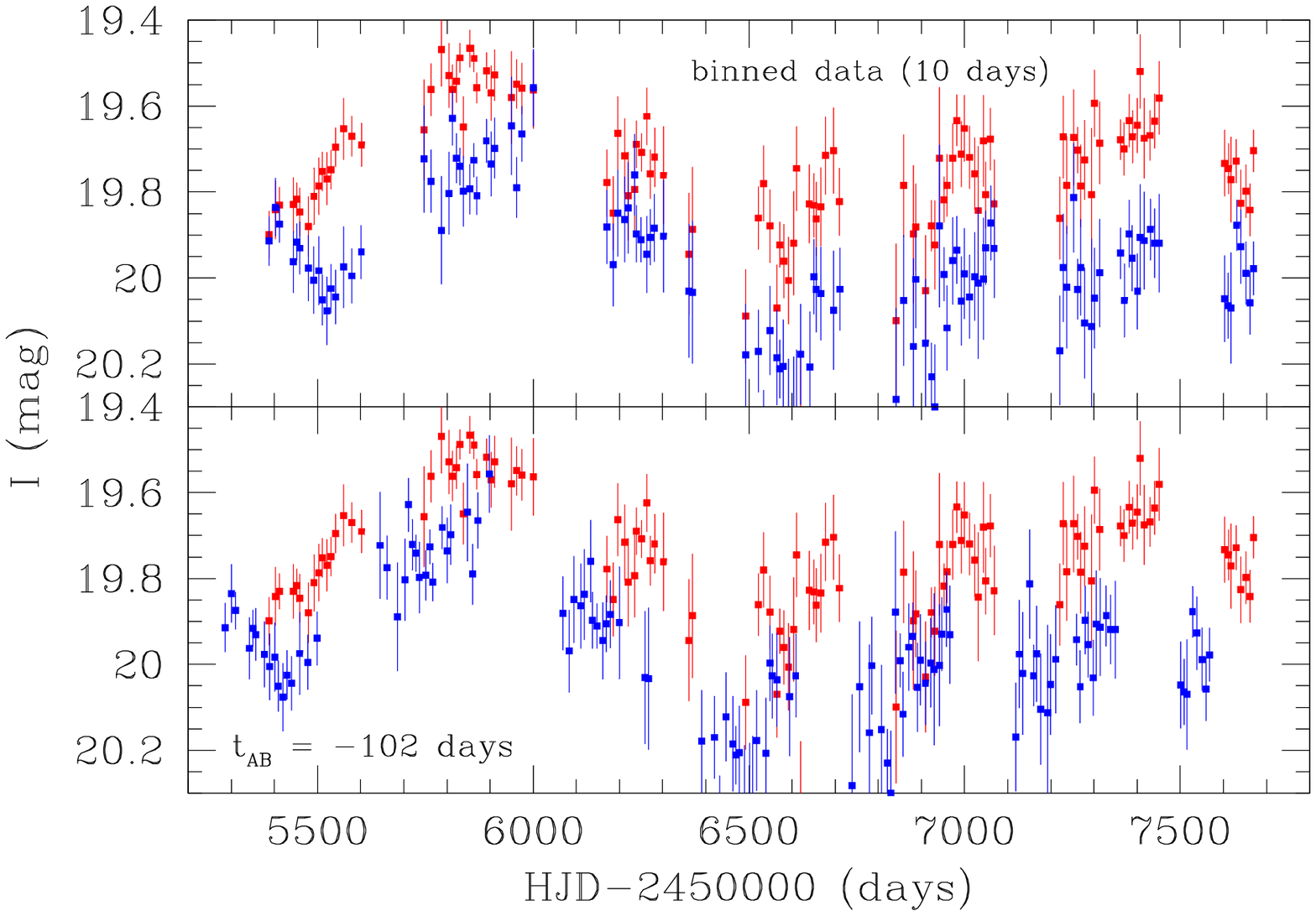}\\
    \vspace{0.1cm}\\
    \includegraphics[width=\columnwidth]{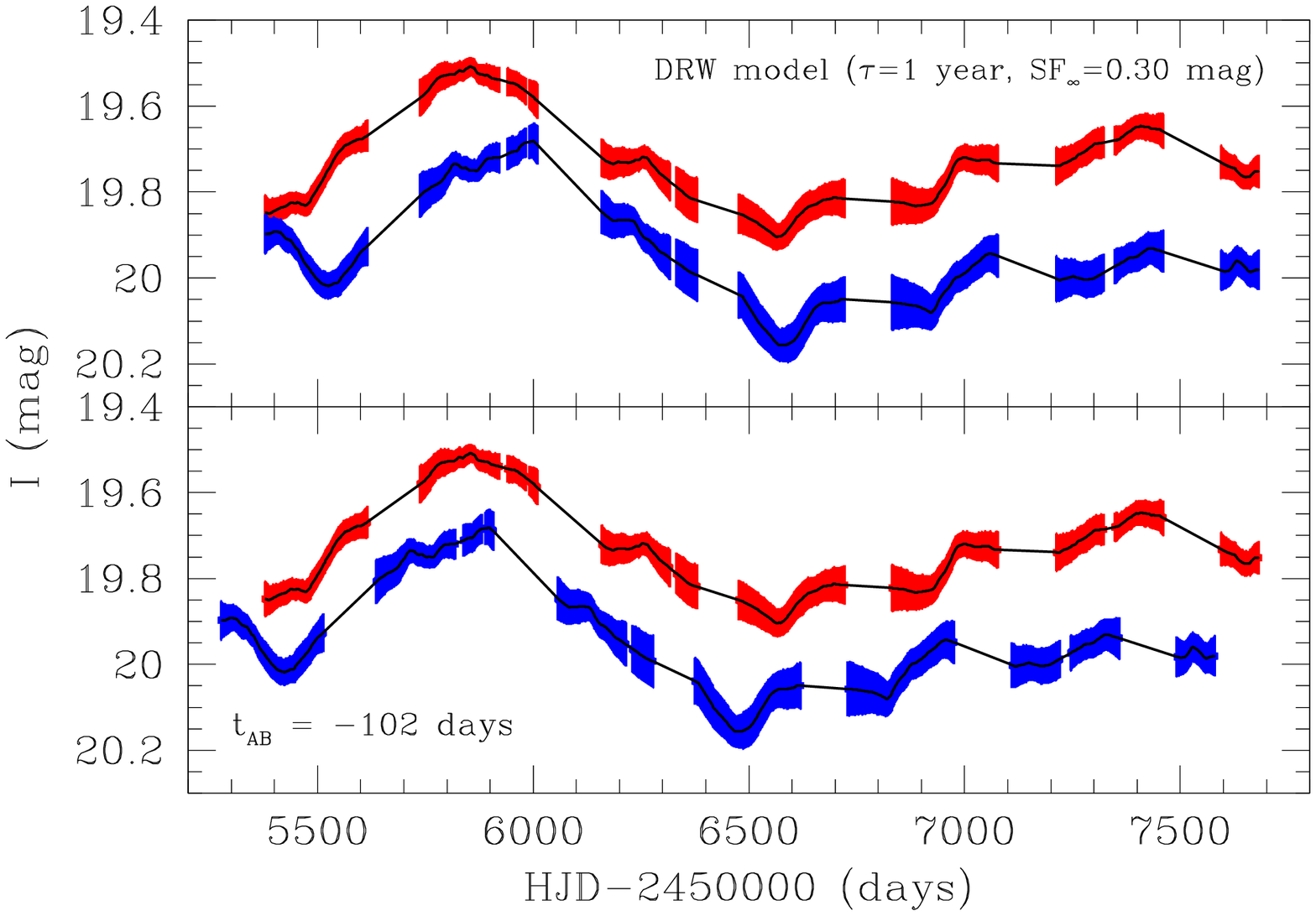}
    \caption{Top panels: 10-days-binned light curves of the two images A (red) and B (blue) 
    of the lensed quasar. The upper panel shows the data as observed, while the bottom one 
    is for image B shifted by $-102$ days (see Section~\ref{sec:timedelay}).
    Bottom panels: We present the DRW models for both images, where the data are inter- and extrapolated
    to within 10 days of each original data point. Similarly to the upper panels, we show the best model 
    light curves as observed (top) and with the image B shifted by $-102$ days (bottom).}
    \label{fig:LCs}
\end{figure}

Time delay(s) are estimated from light curves of the two (or more) images of the quasar. The necessary condition is that all light curves cover the same variability pattern intrinsic to the quasar. This usually means a need for about several years-long, well-sampled light curves having a good signal-to-noise (i.e., sufficiently bright images), because the typical AGN variability amplitude is small, a few tenths of a magnitude, on a time scale of months to years (e.g., \citealt{2010ApJ...721.1014M,2016ApJ...826..118K}). If the separation between the image(s) and the galaxy is small, or simply the galaxy is large (as in the case of QSO 2237$+$0305; e.g.,  \citealt{1985AJ.....90..691H,2006AcA....56..293U}), then one or more images may undergo secondary microlensing by stars of the lensing galaxy. This happens rather frequently and is the main reason of troubles with correctly estimating time delays (e.g., \citealt{1998ApJ...492...74G,2017arXiv170701908T}) and the image flux ratios (e.g., \citealt{2002ApJ...575..650W}), while on the other hand enabling the measurement of the intrinsic properties of the quasar accretion disk (its temperature profile and size, e.g., \citealt{2004ApJ...605...58K,2010ApJ...709..278D,2012ApJ...756...52M,2013ApJ...769...53M}).

\begin{table}
	\centering
	\caption{Time delays obtained using the \protect\cite{2006ApJ...640...47K} method (see Section~\ref{sec:timedelay} for explanations).\label{tab:Koch}}
	\begin{tabular}{lccr} 
		\hline
		$N_{\rm src}$ & $N_\mu$  & parameter  & value \\
		\hline
		10 & 1 & best         & $t_{\rm AB}=-89$ days \\
		10 & 1 & 1$\sigma$    & $-106 < t_{\rm AB} < -72$ days \\
		10 & 1 & 2$\sigma$    & $-124 < t_{\rm AB} < -56$ days \\
		10 & 1 & 3$\sigma$    & $-142 < t_{\rm AB} < -41$ days \\
		10 & 1 & $\chi^2$/dof & 222.6/219 \\
		\hline
		20 & 1 & best         & $t_{\rm AB}=-84$ days \\
		20 & 1 & 1$\sigma$    & $ -97 < t_{\rm AB} < -72$ days \\
		20 & 1 & 2$\sigma$    & $-111 < t_{\rm AB} < -60$ days \\
		20 & 1 & 3$\sigma$    & $-126 < t_{\rm AB} < -48$ days \\
		20 & 1 & $\chi^2$/dof & 185.6/209 \\
		\hline
		30 & 1 & best         & $t_{\rm AB}=-84$ days \\
		30 & 1 & 1$\sigma$    & $ -99 < t_{\rm AB} < -71$ days \\
		30 & 1 & 2$\sigma$    & $-116 < t_{\rm AB} < -49$ days \\
		30 & 1 & 3$\sigma$    & $-132 < t_{\rm AB} < -32$ days \\
		30 & 1 & $\chi^2$/dof & 167.1/199 \\
		\hline
	\end{tabular}
\end{table}

A straightforward method to measure time delay(s) is the cross-correlation (CC) of the two (or more) light curves, however because the data points are rarely regularly sampled, one needs to interpolate (or sometimes extrapolate) one of the light curves to match the points of the first light curve. The exact choice of inter/extrapolating/modelling method is tricky -- one can simply use the spline/straight line or similar method of interpolation or one can try to model the quasar variability as the damped random walk (DRW; e.g., \citealt{2009ApJ...698..895K,2010ApJ...708..927K,2010ApJ...721.1014M}) or even more sophisticated continuous-time autoregressive moving average (CARMA; e.g., \citealt{2014ApJ...788...33K}) models and then simultaneously search for the time delays.

\begin{figure}
\vspace{-1cm}
	\includegraphics[width=\columnwidth]{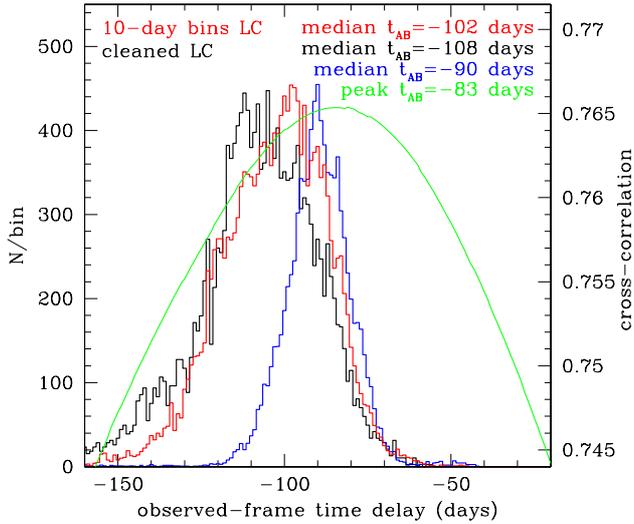}
    \vspace{-0.5cm}
    \caption{The observed-frame time delay density probability obtained using {\sc Javelin} on the cleaned (black) and 10-day binned (red) OGLE light curves is shown.
	The cleaned light curves have the 90\% time delay limits spanning $-151<t_{\rm AB}<-75$~days with the median of $t_{\rm AB}=-108$~days, while 
	the binned light curves have $-132<t_{\rm AB}<-76$~days and $t_{\rm AB}=-102$~days, respectively. The blue line is the resulting time delay distribution from our {\sc python} code, that returned the median time delay of $t_{\rm AB}=-90$~days, where the 90\% confidence limits are $-119<t_{\rm AB}<-57$~days.  
In green (right y-axis), we show the cross-correlation of the 10-day binned data interpolated with DRW model, that broadly peaks at $t_{\rm AB}=-83$~days.}
    \label{fig:DT}
\end{figure}

In this paper, to increase the signal-to-noise ratio in the light curves, we combine the data into 10-day contiguous bins, and then whenever necessary interpolate such light curves using DRW model (Figure~\ref{fig:LCs}). From a structure function analysis, we know that the asymptotic variability is ${\rm SF}_\infty=0.30$~mag and that the structure function power-law slope is $\gamma\approx0.5$, as required by DRW model. The DRW time scale was fixed to 1 year rest-frame (\citealt{2016ApJ...826..118K}). We cross-correlated such a DRW-interpolated light curve to obtain a weakly constrained time delay of $t_\mathrm{AB}\approx-83$ days and allowing for a wide range of negative values (Figure~\ref{fig:DT}), where the image A leads the image B what is in line with expectations from the geometry as image A is farther from the galaxy and it should always lead.

\cite{2006ApJ...640...47K} provide a method of estimating the time delay between the images of the quasar, where the intrinsic variations of the source can be approximated as a Legendre series of order ${\rm N_{src}}$ and are present in all images shifted by the time delay, while the secondary variations due to microlensing are approximated by another Legendre series of order ${\rm N_\mu}$. The orders of both series are typically set by using the F-test. In Table~\ref{tab:Koch}, we present 
the results for three low-order variability polynomials ${\rm N_{src}}=10$, 20, and 30, and the lowest order microlensing polynomial with ${\rm N_\mu}=1$. The time delays are typically within the 1$\sigma$ range of $-100<t_{\rm AB}<-70$ days.

The {\sc Javelin} software is designed to search for time delays between two or more time series, where the time series is intrinsically modelled as the DRW model (\citealt{2009ApJ...698..895K,2010ApJ...708..927K,2010ApJ...721.1014M}). However, it does not simultaneously model any additional variability in each light curves due to microlensing. We used {\sc Javelin} to search for time delays for both full (but cleaned of bad points) and 10-day-binned light curves. 
The time delay density probability distributions are presented in Figure~\ref{fig:DT}. The median time delays are
$t_{\rm AB}=-108$~days and $t_{\rm AB}=-102$~days, respectively, while the 90\% confidence limits are $-151<t_{\rm AB}<-75$~days and $-132<t_{\rm AB}<-76$~days. As a cross-check, we also used 
our own {\sc python} DRW-modelling time-delay searching code. The results are comparable, although the time delay is a little shorter, with the median $t_{\rm AB}=-90$~days and the 90\% confidence limits of $-119<t_{\rm AB}<-57$~days.

The cross-correlation, the Kochanek method, and the {\sc python} code seem to prefer slightly shorter, but not disjoint, time delays to the ones derived from the {\sc Javelin} code. For further discussions, throughout this paper we will adopt the following median time delay of $t_{\rm AB}=-102$~days and the 90\% confidence level of $-132<t_{\rm AB}<-76$~days, which are the results from the {\sc Javelin} modelling of the 10-day binned data.

\section{Discussion}
\label{sec:discussion}

\subsection{Modelling the Lensing System}

\begin{figure}
	\includegraphics[width=\columnwidth]{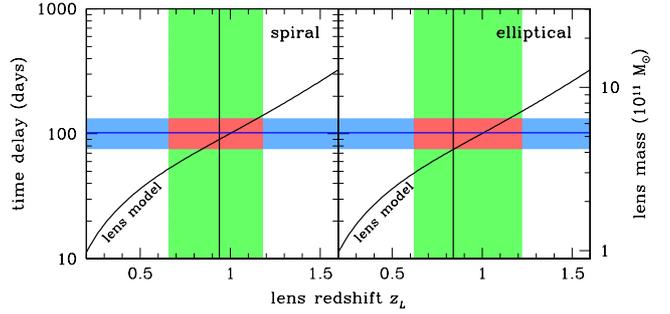}
    \caption{The time delay (left y-axis) and the approximate lens mass (right y-axis) as a function of redshift. The inclined line shows the prediction based on the simple SIE lens model. The horizontal blue line indicates the time delay of $|t_{\rm AB}|=102$ days, while the light blue region represents 90\% confidence levels. The two vertical lines mark the best-fitted SED redshifts for the spiral (left panel; $z_L=0.94$) and elliptical galaxy (right panel; $z_L=0.84$), when the source redshift is fixed to the spectroscopically measured value ($z_S=2.16$) in the SED fitting. The green areas mark the 90\% confidence levels for these redshifts, while in red, we mark the overlapping 90\% confidence levels for both the time delay and redshifts.
The best estimate for the lens galaxy mass is $\sim$$5\times10^{11}$~$M_\odot$ within the Einstein radius.}
    \label{fig:zettm}
\end{figure}

The available data and their uncertainties limit our ability to fully model the lensing system. As an approximation, we modelled the lens by adopting the singular isothermal ellipsoid (SIE) model for the lens mass density profile (\citealt{1994A&A...284..285K}), and using the {\sc gravlens} software (\citealt{2011ascl.soft02003K}).
We obtained the Einstein radius of $R_E=0.78 \pm 0.01$ arcsec, axis ratio $0.91 \pm 0.05$ and position angle $90 \pm 20$ (in degrees measured east of north) from exploring the parameter space with MCMC sampling. The basic SIE model leads to the discrepancy in fitted values of axis ratio between mass and light profile but one has to remember that the external shear was not taken into account and/or the SIE model may be too simplistic. For the measured time delays, the {\sc gravlens} modelling results point to the lens redshift (used here as a free parameter and so sampled for) and the lens mass. The lens redshift is found to be $0.83-1.12$ (90\% CL) and the lens mass within the Einstein radius $4-6\times 10^{11} \mathrm{M_\odot}$ (see Figure~\ref{fig:zettm}).
The estimated lens redshift from the SED modelling is in excellent agreement with our prediction from the lens model, however, the uncertainties are large.

The best-estimated total lens galaxy mass implies a stellar mass of $2\times10^{11} \mathrm{M_\odot}$ assuming a typical stellar-to-total mass ratio within the Einstein radius of $\sim0.5$ (e.g., \citealt{2010ApJ...724..511A}). This number can be compared against the 
galaxy mass distributions for late-type (spiral and irregular) and
early-type (elliptical and lenticular) galaxies from SDSS. Our derived mass is by far more likely found
in elliptical galaxies than in spirals (\citealt{2014MNRAS.440..889S}). Another confirmation comes from the SED modelling, where
we estimate the stellar mass for the lens being elliptical or spiral galaxy with $(2.5 \pm 0.5) \times 10^{11} \mathrm{M_\odot}$ and $(3.3 \pm 0.7) \times 10^{11} \mathrm{M_\odot}$, respectively. Both values broadly agree with the expectation from the lens model.

\subsection{Detectability of lenses in OGLE}

According to predictions presented by \cite{2010MNRAS.405.2579O}, there should be   
1 gravitational lens for $I<19$ mag, 3 lenses for $I<20$ mag, 8 lenses for $I<21$ mag, 
and 20 lenses (both doubles and quads) for $I<21.5$ mag (the detection limit in OGLE)
within the area of about 670 square degrees around the Magellanic Clouds. 
We adopted the typical quasar colour of $(i-I)\approx0.4$~mag (Figure 9 in \citealt{2012ApJ...746...27K}) 
to convert the \cite{2010MNRAS.405.2579O} $i$-band magnitudes to $I$-band.
Our candidate selection was partly based, however, on a requirement of detecting quasar variability. 
\cite{2016ApJ...826..118K} showed that the rms of the optical variability of quasars reaches about 0.25 mag, provided the light curves are sufficiently long.
The rms of the photometric noise in OGLE-IV is 0.06 mag for $I=19$ mag, 0.13 mag for $I=20$ mag, and 0.30 mag for objects with $I=21$ mag (\citealt{2015AcA....65....1U}).
So the expected number of found lensing systems will be actually lower due to the fact that only a fraction of 
quasars are sufficiently variable around $I \approx 20$ mag to exceed the photometric noise in OGLE. 
Ignoring other factors, such as the high stellar crowding, we should in principle detect fewer than five gravitational lenses.
This number still could be lower in our case, because a fraction of the analysed area has been only observed for three years at the time of this search,
so a fraction of quasars show lower rms variability. This implies  that our detection efficiency is at least 20\% ($>$1/5).

\section{Conclusions}

We presented the results from our search for lensed quasars in the OGLE Variability Survey's data in the extended area of the Magellanic Clouds. 
Our first candidate, that we confirmed spectroscopically, is a lensed quasar at a redshift of $z_S=2.16$, with two quasar images separated by 1.5\arcsec  on the sky. 
After subtracting the two lensed images from a deep OGLE frame, we detected the lensing galaxy with $I\approx21.5$~mag. 
From the SED modelling spanning 0.2--24 microns and having the fixed source redshift, we found that the lensing galaxy could be both a spiral or an elliptical at the redshift of $z_L\approx0.9\pm0.2$,
although an elliptical galaxy is preferred. This redshift is also expected from the simplest SIE lens model with the measured time delay of $-132<t_{\rm AB}<-76$~days (90\% CL). 

If the lens was a spiral this would raise a potential question on the amount of the expected extinction to be present in the SED of the quasar images (but undetected), because the lines of sight pass approximately 8 kpc (for image A) and 3 kpc (for B) from the galaxy lens center. For a very massive spiral with the stellar mass of $3 \times 10^{11}~M_\odot$ this would likely mean passing through the spiral arms containing dust (unless it is edge on and has a favourable position angle, the latter being the case here as the PA is roughly perpendicular to the line joining the images). However, the lensing mass is nearly circular which is inconsistent with an edge-on disk. Furthermore, our derived stellar mass is typically seen in elliptical galaxies and is generally too high, by roughly an order of magnitude, for spirals (\citealt{2014MNRAS.440..889S}). 

To obtain the time delay of $-132<t_{\rm AB}<-76$~days (90\% CL), we used several available methods that included the DRW light curve modelling with {\sc Javelin}, but also with our
own {\sc python} code, the \cite{2006ApJ...640...47K} method using the Legendre polynomials, and also cross-correlation.
All these methods provided consistent results, although the {\sc Javelin} software preferred slightly longer time delays.

We proposed another detection route of gravitationally lensed systems, first through the ``red'' mid-IR WISE colours, typical for quasars, 
and then including the identification of point source pairs in optical data showing similar variability patterns. 
Our search method does not require any prior spectroscopic observations.


\section*{Acknowledgements}
We are grateful to the anonymous referee for providing us with comments that improved this manuscript. 
ZKR acknowledges support from European Research Council Consolidator Grant 647208.
SK acknowledges the financial support of the Polish National Science Centre through the OPUS grant number 2014/15/B/ST9/00093.
TA and YA acknowledge support by proyecto FONDECYT 11130630 and by the Ministry for the Economy, Development, and Tourism's Programa Inicativa Cient\'{i}fica Milenio through grant IC 12009, awarded to The Millennium Institute of Astrophysics (MAS).
OGLE project has received funding from the Polish National Science Centre, grant MAESTRO 2014/14/A/ST9/00121 to AU.
Part of the funding for GROND (both hardware as well as personnel) was generously granted from the Leibniz-Prize to Prof. G. Hasinger (DFG grant HA 1850/28-1).

This publication makes use of data products from the Wide-field Infrared Survey Explorer, which is a joint project of the University of California, Los Angeles, and the Jet Propulsion Laboratory/California Institute of Technology, funded by the National Aeronautics and Space Administration. 



\bibliographystyle{mnras}


\appendix
\section{Other observed candidates}
\counterwithin{figure}{section}

\begin{figure*}
    \includegraphics[width=0.48\columnwidth]{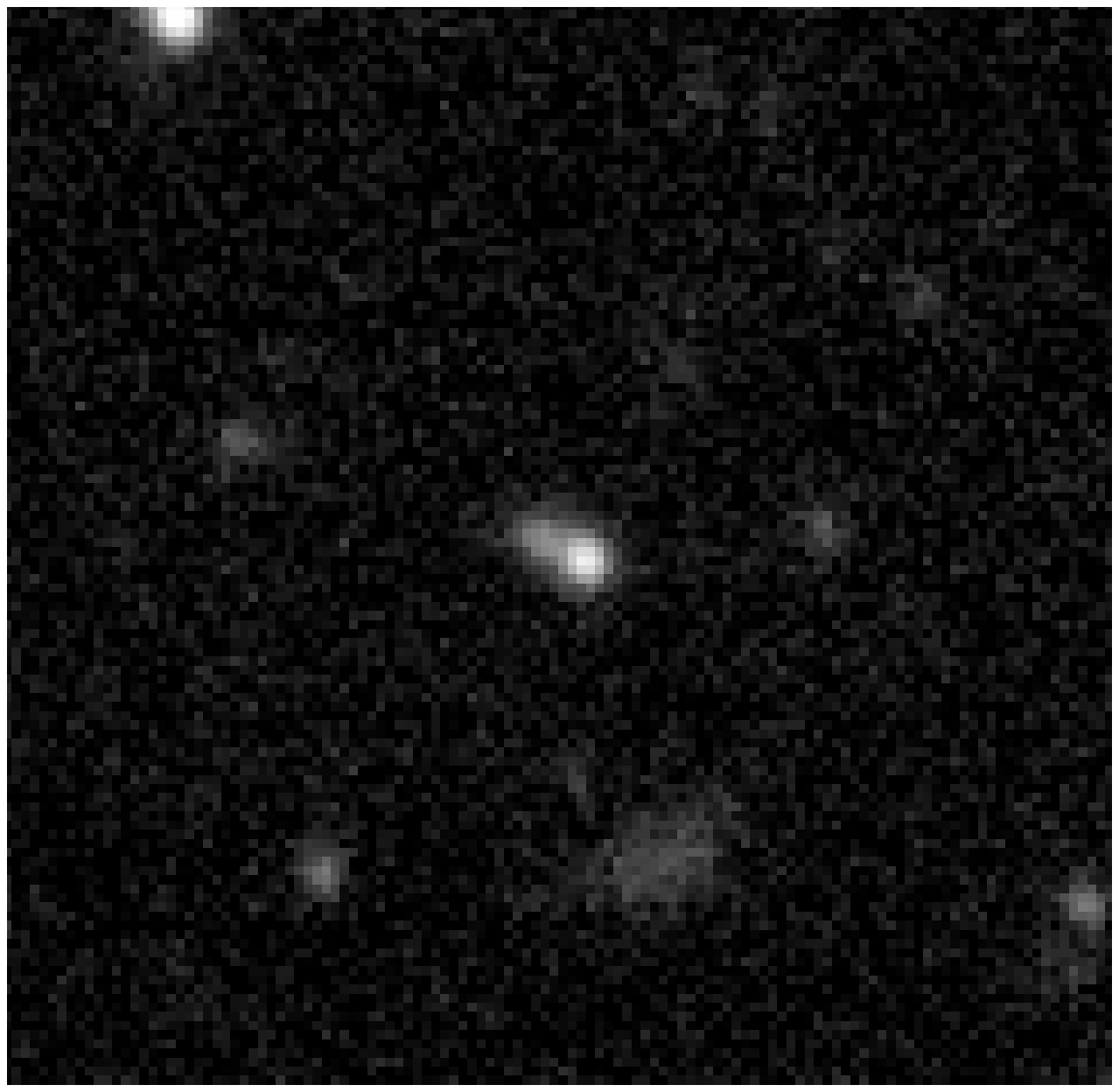} \hspace{0.2cm}
    \includegraphics[width=0.5\columnwidth]{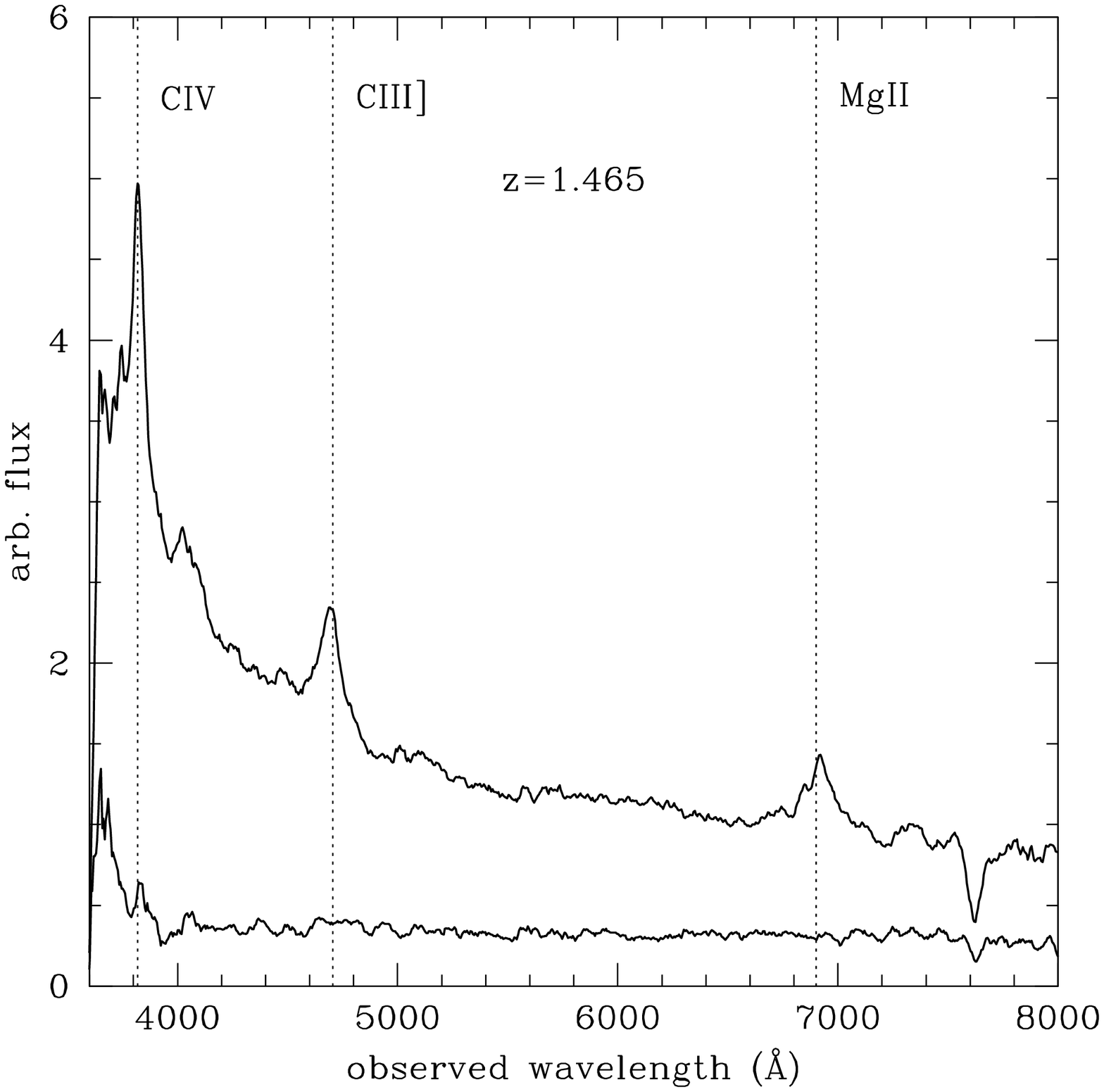} \hspace{0.2cm}
    \includegraphics[width=0.5\columnwidth]{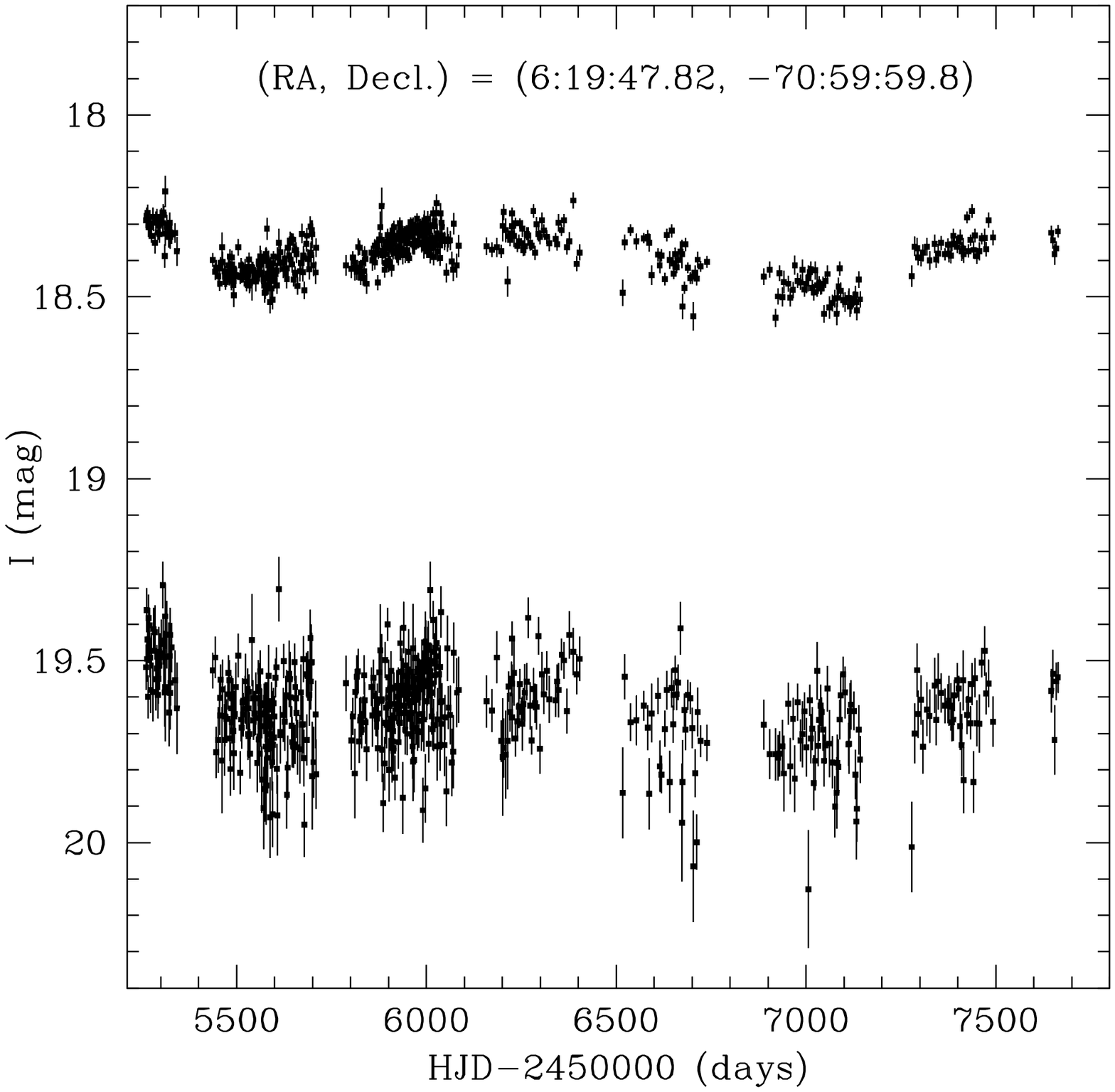}\\
    \vspace{0.2cm}
    \includegraphics[width=0.48\columnwidth]{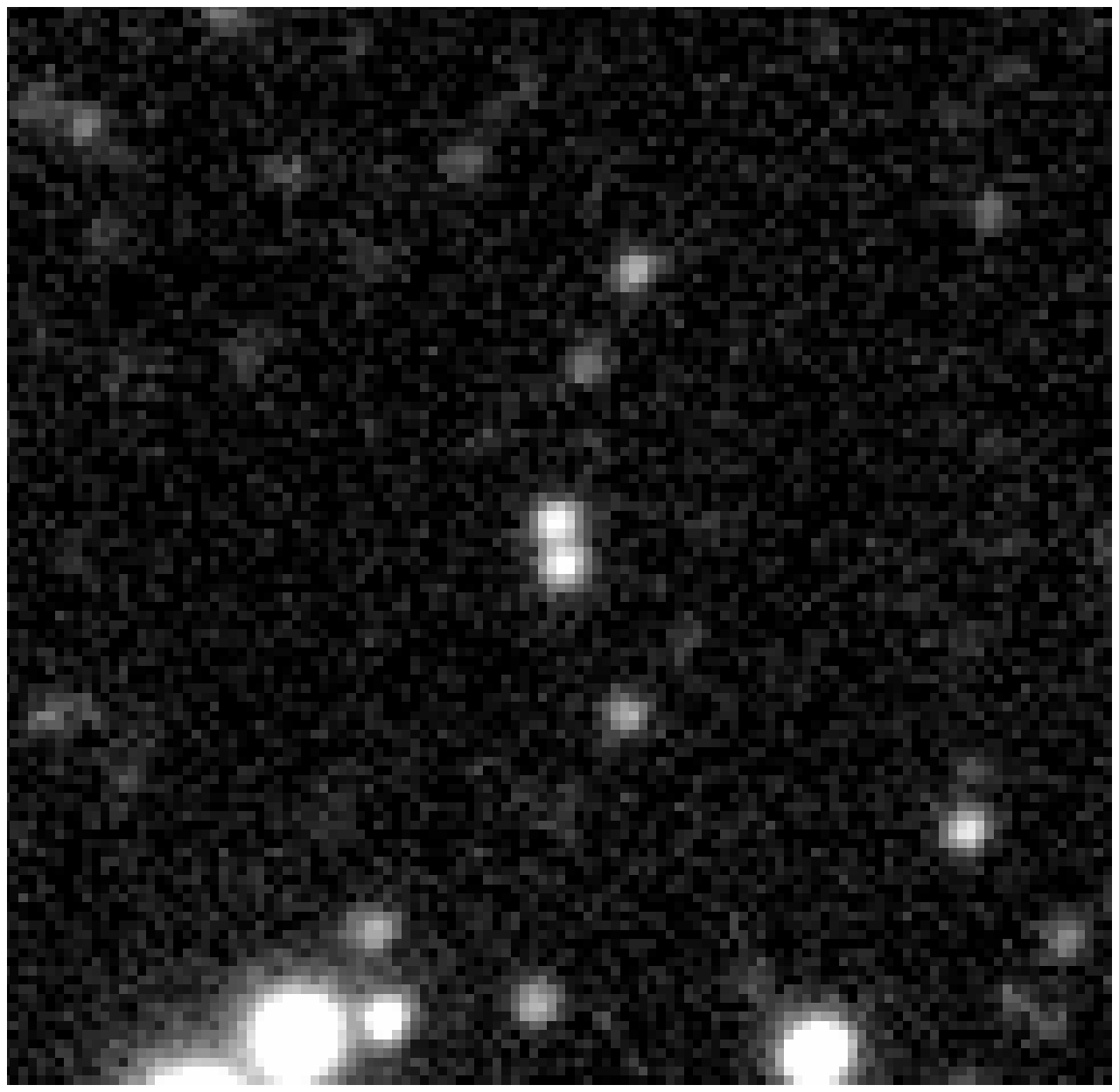} \hspace{0.2cm}
    \includegraphics[width=0.5\columnwidth]{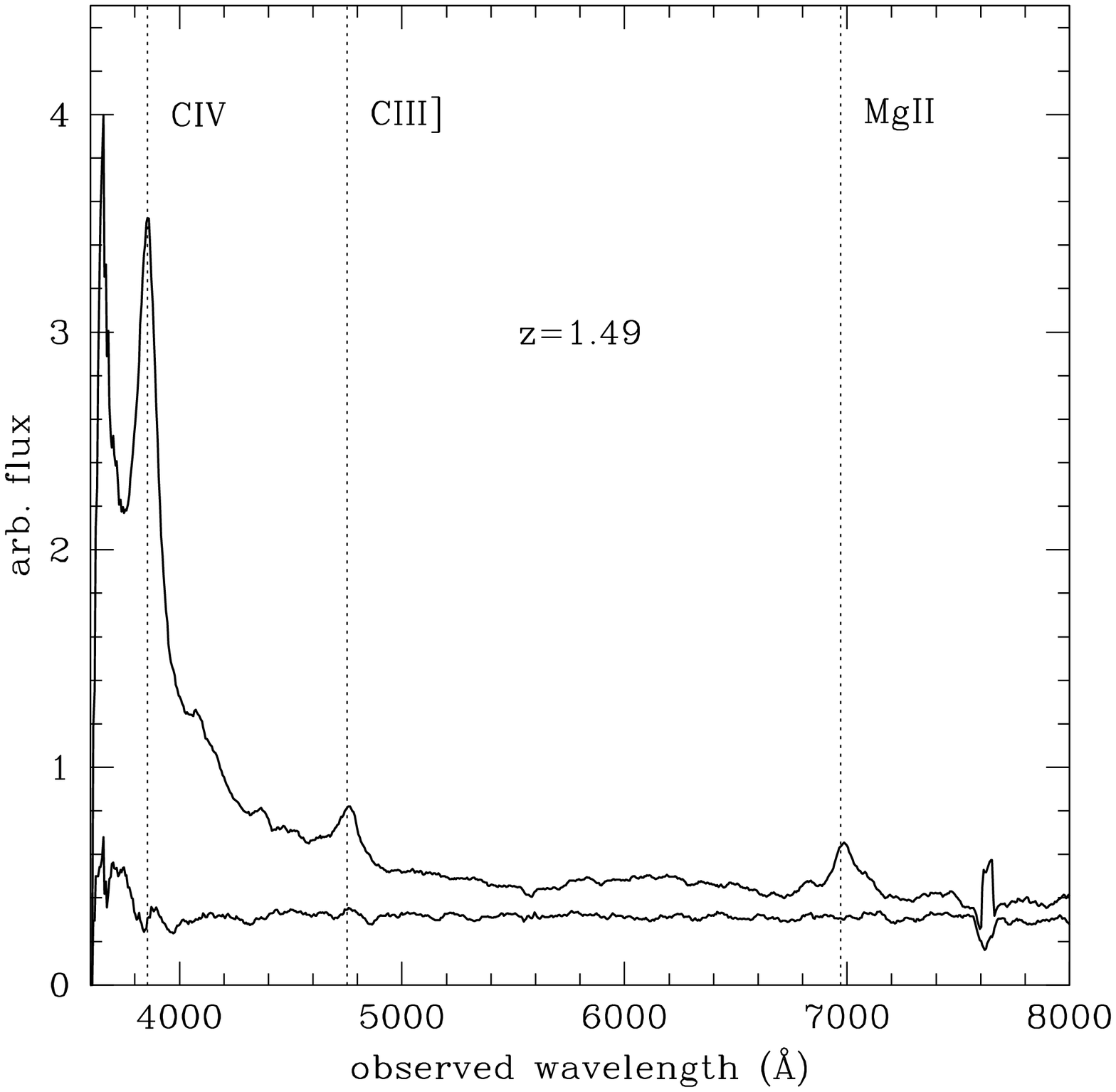} \hspace{0.2cm}
    \includegraphics[width=0.5\columnwidth]{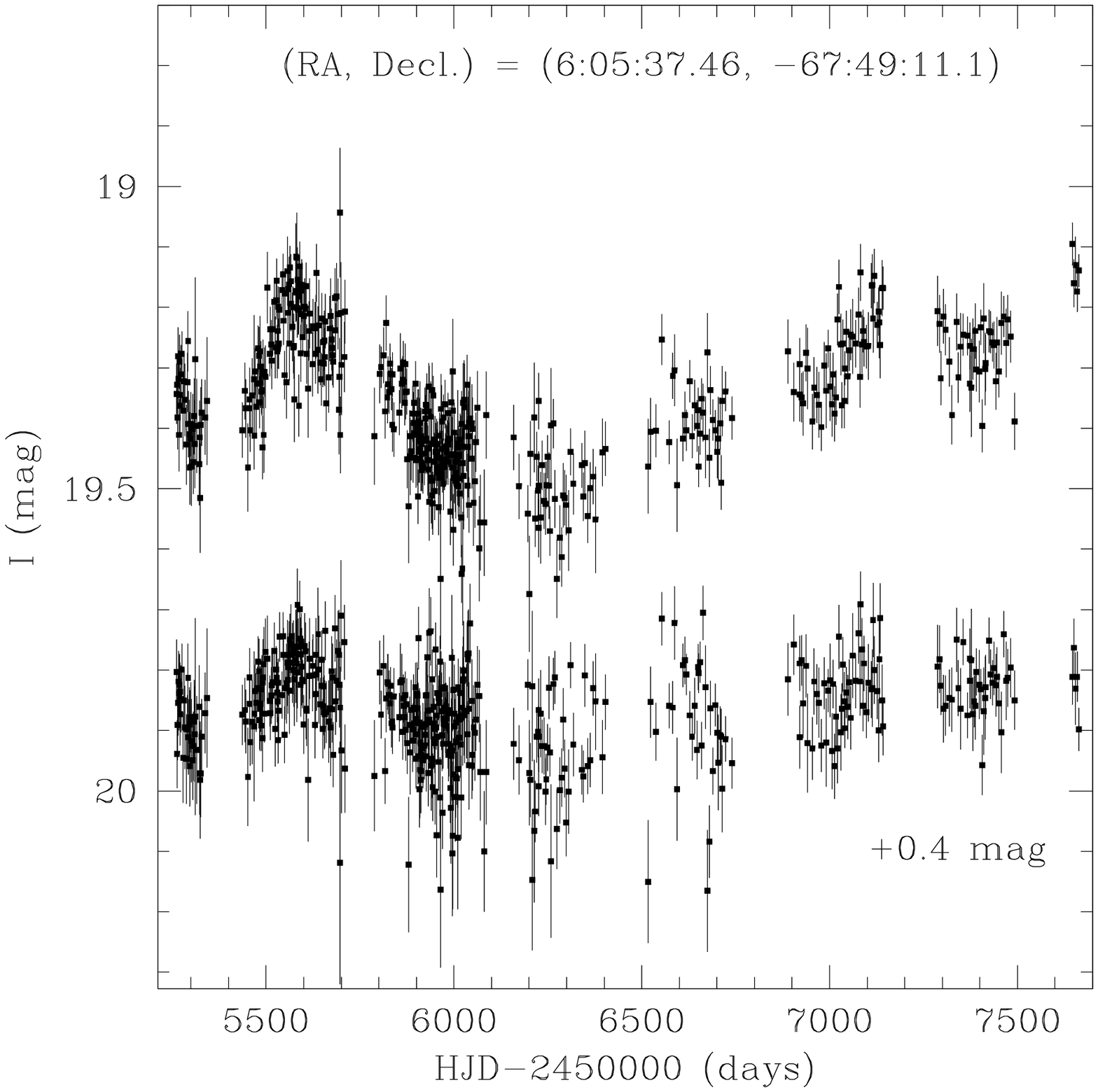}
    \caption{Two spectroscopically observed candidates that turned out to be regular quasars with a nearby interloper. In the left panel, we present 1\arcmin\ by 1\arcmin\ finding charts centred at these sources, in the middle panels we show the NTT spectra with the main quasar emission lines marked as vertical lines, and in the right panels we present the OGLE-IV light curves. The best estimate {\sc Javelin} time delays are $-2$ days (top) and 15 days (bottom). }
    \label{fig:contam}
\end{figure*}

Here we present basic information on the two spectroscopically observed lensed quasar candidates which turned out to be unlensed quasar+star pairs. In Fig. \ref{fig:contam} we present the finding charts, the NTT spectra and the OGLE light curves in $I$-band. Both candidates were observed on 2016 December 5 with the EFOSC2. Both quasars show common AGN broad emission lines at the redshift of $z=1.465$ and $z=1.49$. The spectra of nearby interlopers are also presented for comparison.   

\bsp	
\label{lastpage}
\end{document}